\documentclass{aastex631}
\usepackage[flushleft]{threeparttable}
\usepackage{natbib}
\usepackage{url}
\usepackage{placeins}
\def\UrlBreaks{\do\/\do-}

\newcommand{\mycomment}[1]{}

\begin{document}


\shortauthors{Keto \& Watters}
\shorttitle{Moving Objects in EOS Images}

\title{Detection of Moving Objects in Earth Observation Satellite Images: Verification}


\author{Eric Keto}
\affiliation{Harvard University, Institute for Theory and Computation \\
60 Garden Street,
Cambridge, MA USA}

\author{Wesley Andr\'{e}s Watters}
\affiliation{Wellesley College, Whitin Observatory \\
106 Central St.,
Wellesley, MA 02481, USA}




\begin{abstract}
In multi-spectral images made by Earth observation satellites that use push-broom scanning, 
such as those operated by Planet Labs Corp., 
moving objects can be identified by the appearance of the object at a different locations in each spectral band.
The apparent velocity can be measured if the relative acquisition time between images in different spectral bands is 
known to millisecond accuracy. The images in the Planet Labs archive are mosaics of individual
exposures acquired at different times. Thus there is not a unique acquisition time for 
each spectral band.
In an earlier paper, we proposed a method to determine the relative acquisition times  
from the information in the images themselves. 
High altitude balloons provide excellent
targets to test our proposed method because of their high apparent velocity 
due to 
the orbital velocity of the satellite and geometric parallax in images
aligned to the level of the ground.
We use images of the Chinese balloon that crossed the US in February, 2024 as well as images
of an identical balloon over Colombia to test our method. 
Our proposed method appears to be successful and allows the measurement of the apparent velocity 
of moving objects from the information available in the archive.
\end{abstract}


\section{Introduction}
 
The satellites of Planet Labs Corp. (PLC)  image most of the entire land area of the Earth with a daily revisit rate. 
The PLC archive of daily images extends  
back to 2020. While the satellites are not intended to provide motion detection as in a high frame-rate video, the arrangement
of the filters to provide multi-spectral imaging in combination with the orbital motion of the satellite (push-broom scanning)
provides a short video with the number of frames equal to the number of spectral bands. 
For example, in an RGB  composite visual image, an object moving with sufficient apparent velocity
appears as three individual objects with red, green, and blue colors. Images from PLC's most recent
generation of satellites, named SuperDoves, show this effect in eight spectral bands and provide a video of an an area
equal the field of view with a duration equal to the time required for all spectral bands to cross the field of view, 
$\sim 3.2$ s duration. This allows a repurposing of the archived images to detect objects moving at speeds from about ten m$^{-1}$
to several hundred m s$^{-1}$ over most of the Earth during the times that the images were acquired, typically once per day around
noon local time.

The analysis of the images to detect motion is trivial if the acquisition times of individual exposures are known. However, the images in
the Planet Labs archive are mosaics of smaller area individual exposures and the acquisition times of the individual regions that
make up the archived mosaic are not available in the archive. The mosaicing pattern is different for each composite image because the
exposure or frame rate of the camera is asynchronous with the time required for the satellite to cross the field of view. Furthermore, the mosaicing algorithm as well as the details of the design of the satellites are considered
proprietary technology by Planet Labs.

In an earlier article \citep{KW2023}, we proposed that it is nonetheless possible to determine enough information from the archived images
to estimate the relative acquisition times between the different spectral bands in each region of the mosaic
containing a moving object and
then measure the velocities and altitudes of moving objects, subject to a well-known ambiguity between the velocity
in a direction parallel to the orbital track of the satellite and the altitude of the object. 
The principal assumption of our proposed method is that the camera on each
satellite operates at fixed frame rate during the $\sim 3.2$ s time required to acquire enough images to make one mosaic.

In this article, we test this method using images of high-altitude balloons.  High-altitude balloons are advantageous targets
because the the motion of the balloon itself can be ignored in the analysis. The wind speed in the lower stratosphere, $\sim 20$ km altitude,
is typically $< 10$ ms$^{-1}$ \citep{Limpinsel2018}, and slow with respect to the apparent velocity due to parallax at altitude, $\sim 300$ ms$^{-1}$.
Additionally, the balloons are
incapable of significant acceleration within the $\sim 3.2$ s acquisition time that otherwise might confuse the analysis.
We find the method successful. In the two examples studied, both with high signal-to-noise and high spatial resolution, 
we estimate
errors less than 2\% on the velocity and 3\% on the altitude from the standard deviation of the individual 
measurements. 

\section{Selection of Targets}

\begin{figure}[!h]
\includegraphics[width=3.0in,trim={0 0 2.0in 6.0in},clip]{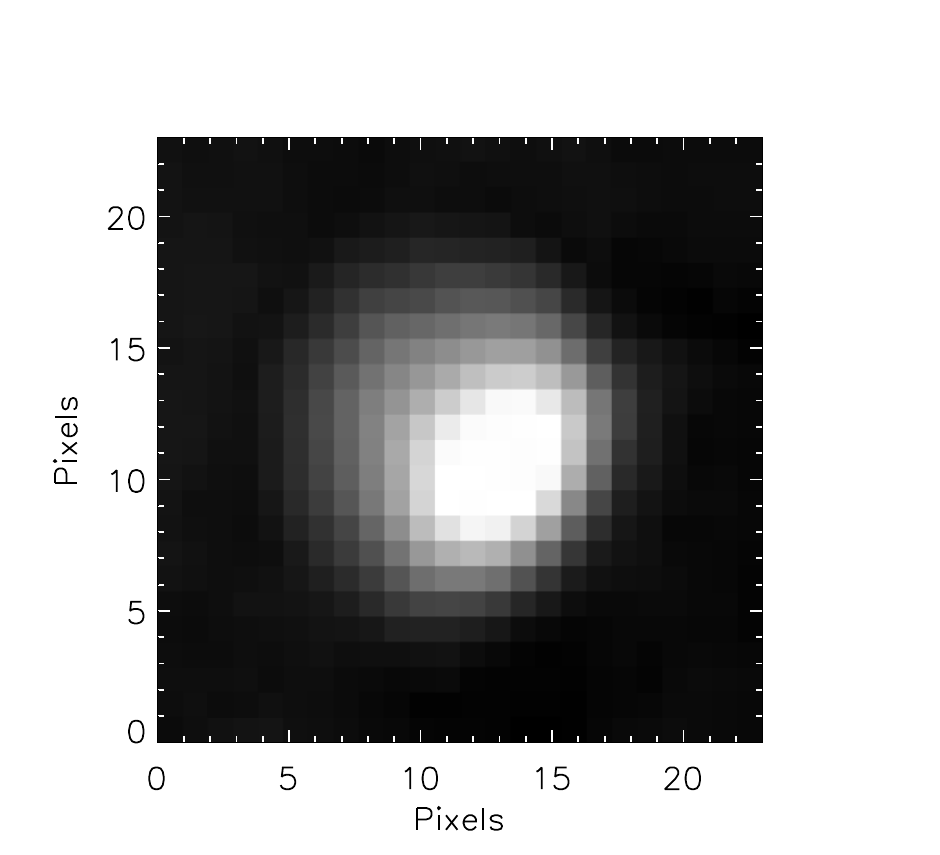}
\caption{
Image of a Chinese balloon over Columbia, near Cartagena, in visual blue, band 0, of the SuperDove satellite 247d. 
The pixel size is (3 m)$^2$. }
\label{template2479}
\end{figure}

There are numerous high altitude weather balloons that would serve as test objects. The passage of a large, high-altitude
Chinese balloon across the continental US, February 3-6, provided an ideal target. The balloon is one of the larger
high-altitude balloons, $\sim 40$m versus $\sim 6$ m for a typical weather balloon. At ten times the resolution limit
of the Planet Labs satellites, good positional accuracy is obtained with high signal-to-noise.
Figure \ref{template2479} shows the appearance of the Chinese balloon over 
British Columbia 
near the US-Canadian border in an image from satellite 2479.
In addition, the publicity from the ensuing political antics played out in 
denials\footnote{The airship is from China. It is a civilian airship used for research, mainly meteorological, purposes. Affected by the Westerlies and with limited self-steering capability, the airship deviated far from its planned course. The Chinese side regrets the unintended entry of the airship into US airspace due to force majeure. The Chinese side will continue communicating with the US side and properly handle this unexpected situation caused by force majeure. --- Ministry of Foreign Affairs of the People's Republic of China \citep{MFAPRC2023} }, 
accusations\footnote{We were able to determine that China has a high-altitude balloon program for intelligence collection that's connected to the People's Liberation Army.  --- John Kirby, coordinator for strategic communications at the National Security Council \citep{Vergun2023}},
and attempts at 
secrecy\footnote{Reporter: Is the position of the balloon classified? \\
Pentagon Press Secretary Air Force Brig. Gen. Pat Ryder: What we're not going to do is get into an hour-by-hour location of the balloon
\ldots\  right now it's over the center of the continental United States. That's about as specific as I'm going to get. \\
Reporter: Does the public not have a right to know if the balloon is over their state? \\
Ryder: The public certainly has the ability to look up in the sky and see where the balloon is. --- from a press conference as reported by the  \citet{WSJ2023} }
helped ensure that the balloon would be quickly located by Planet Labs with help from one of
its partner companies, Synthetaic. 
While searching, this collaboration also discovered a second balloon flying
near Cartagena, Colombia on February 3, 2023 with characteristics identical to the Chinese balloon over the US.

We are grateful to Planet Labs for providing the scene identifications (table\ref{imageid}) 
in the
Planet Labs archives (private communication). From the available images, we selected the 
three made by the eight-band SuperDove satellites rather than the four-band Dove satellites because the former provide more 
positions for our analysis.

\begin{table}[!h]
\caption{Planet Labs images with Chinese balloons}
\begin{tabular}{lccl}
\toprule
Approx. location		& Scene ID	& Satellite ID & Satellite Type \\
\colrule
British Columbia	&
20230131\_182637\_88\_2479 & 2479 &
8-band SuperDove \\
South Dakota			&
20230202\_165236\_51\_242d & 242d &
4-band Dove \\
Missouri				&
20230203\_161055\_87\_2439 & 2439 &
8-band SuperDove \\
South Carolina &
20230204\_151350\_28\_2458 & 2458 &
4-band Dove \\
South Carolina &
20230204\_154439\_74\_2495 & 2495 &
4-band Dove \\
Colombia &
20230203\_150645\_50\_247d & 247d &
8-band SuperDove \\
\botrule
\end{tabular}\label{imageid}
\end{table}
\smallskip

\FloatBarrier
\section{Method of Analysis}

\subsubsection{Locating the balloons}
\begin{figure*}[!ht]
\begin{tabular} {ll}
\includegraphics[width=3.5in,trim={0.in 0.in 2.5in 6.0in},clip] {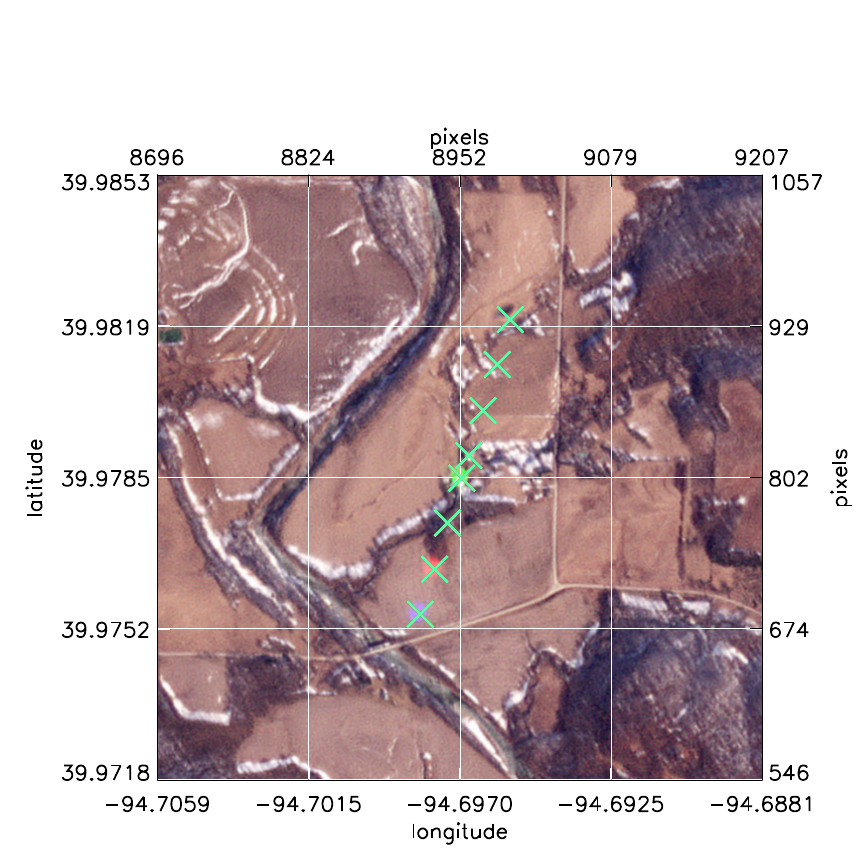} &
\includegraphics[width=3.5in,trim={0.in 0.in 2.5in 6.0in},clip] {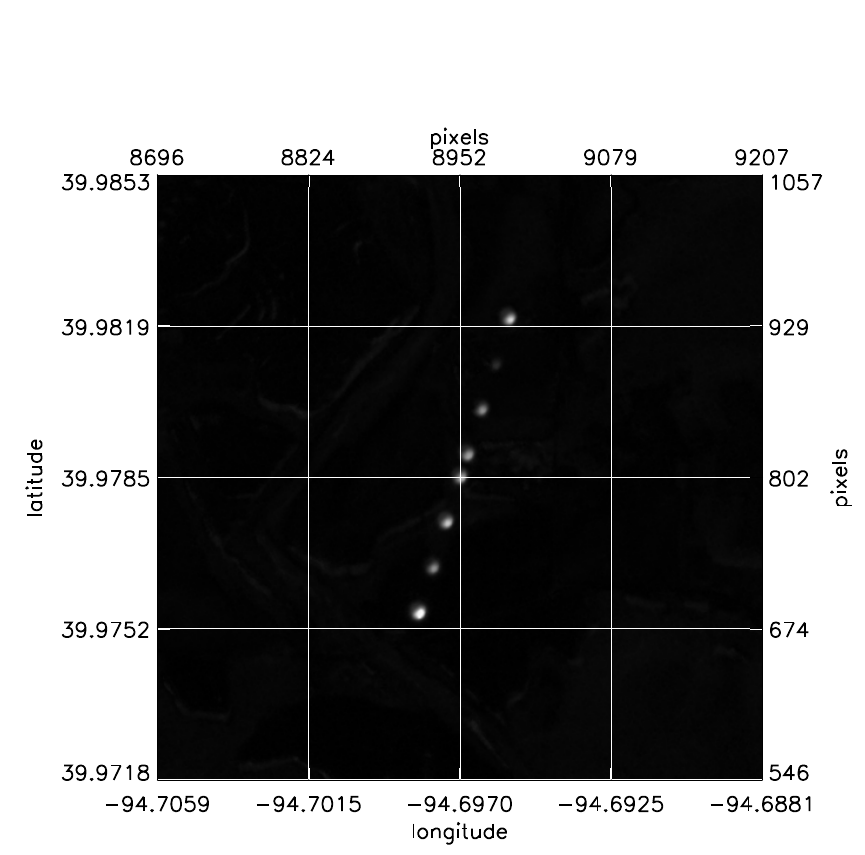} \\
\end{tabular}
\caption{
Positions of the first Chinese balloon over Missouri in the eight bands of SuperDove satellite 2439 against a
3-color visual background ({\it left}) and against the summed absolute values of the seven differenced pairs of
images in the eight spectral bands ({\it right}). The bottom and left axes are labeled by longitude and latitude.
The top and right axes shows the pixel coordinates with respect to the full images in the archive.
On the visual image, the eight positions are marked by crosses.
Only three images of the balloon appear in the red-green-blue visual representation.
The balloon is seen in all eight spectral bands in the composite image of the summed differences ({\it right}). 
In the 8-band image
({\it right}), the balloon in the NIR band, second from the top in both figures, is not as bright because 
the sun is not as bright per unit wavelength in the NIR, and CCDs optimized for visible
wavelengths generally have lower quantum efficiencies in the NIR.
}
\label{positions2439}
\end{figure*}

\begin{figure*}[!ht]
\begin{tabular} {ll}
\includegraphics[width=3.5in,trim={0.in 0.0in 2.5in 6.0in},clip]{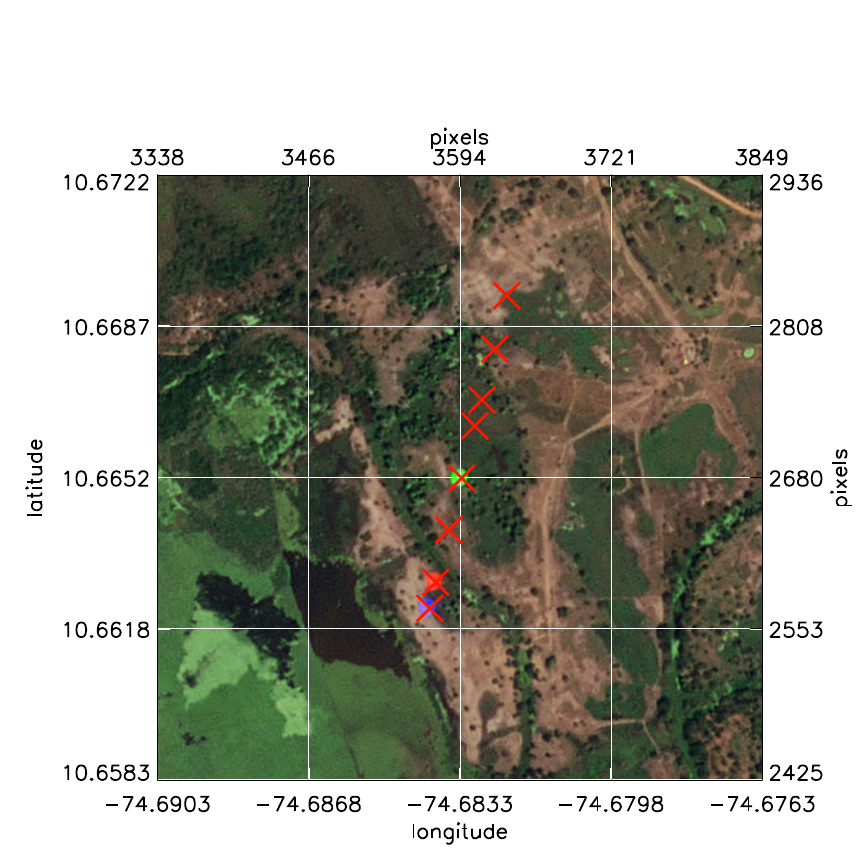} &
\includegraphics[width=3.5in,trim={0.in 0.0in 2.5in 6.0in},clip]{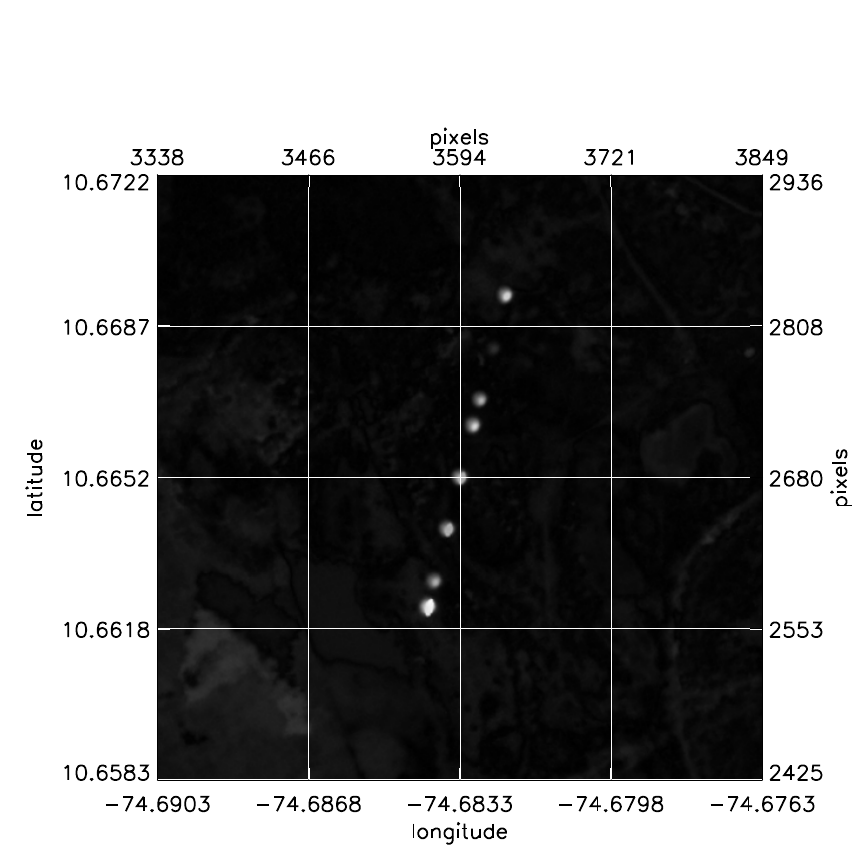} \\
\end{tabular}
\caption{
Positions of the second Chinese balloon over Colombia in the eight bands of SuperDove satellite 247d against a
3-color visual background ({\it left}) and against the summed absolute values of the seven differenced pairs of
images in the eight spectral bands ({\it right}) in the format of figure \ref{positions2439}.
}
\label{positions247d}
\end{figure*}

With their large size and high brightness, the balloons are easily located within the Planet Labs images by differencing
pairs of the eight
spectral bands in spectrally adjacent colors
to show the balloon as a positive and negative pair at different positions against a suppressed background. 
We then use a template matching 
algorithm to locate the positions in each pair.  
Figures \ref{positions2439} and \ref{positions247d} show the locations of the balloons over Missouri and Colombia, respectively, 
in the eight spectral bands of satellites 2439 and 247d (table \ref{imageid}). The locations are shown over two backgrounds. The first 
is a 3-color visual image, and the second is an 8-band image formed from the sum of the absolute values 
of the differences of each pair of images. 

\FloatBarrier
\subsection{Estimating the Relative Timing}\label{Timing}
Our method previously proposed to estimate the apparent velocity of an 
object in Planet Labs images is described in \S 4 of \citet{KW2023}.
The procedure has two steps. In the first step, we estimate the difference in the acquisition times of images in
different spectral bands from the orbit of the satellite and the number of pixels on the sensor. In the second step,
we determine the apparent velocity from the difference of positions of the balloon in each pair of spectral bands
that are adjacent on the spectral filter
and the relative acquisition times
corrected for the mosaicing procedure that is used to combine several camera exposures into an
archived image.

The first step is straightforward. 
From equation 1 of \citet{KW2023}, the time difference
between images of spectral bands whose individual filters are adjacent on the full filter is,
\begin{equation}
\Delta t_{color} = \frac {N_y \mu} {2\pi R_\oplus \omega},
\label{eqn1}\end{equation}
where the width in pixels of one color strip $N_y=663$ is the total length/8 in pixels of the short side of
the rectangular sensor array that is aligned with the direction of the satellite orbit, and
the radius of the Earth, $R_\oplus = 6378$ km. The ground sample distance (GSD) or the length of a square
pixel on the sensor projected on the surface of the Earth is $\mu \sim 4$ m. The exact GSD is returned
in the search results of the Planet Labs archive. The
mean motion, $\omega \sim 15.15$ orbits per day, is available from ephemerides published by Planet Labs with daily two-line element (TLE)
models of the orbit of each satellite. The time difference between images of adjacent spectral bands is then $\Delta t_{color} \sim 0.39$ s.
\begin{figure*}[!ht]
\includegraphics[width=3.5in]{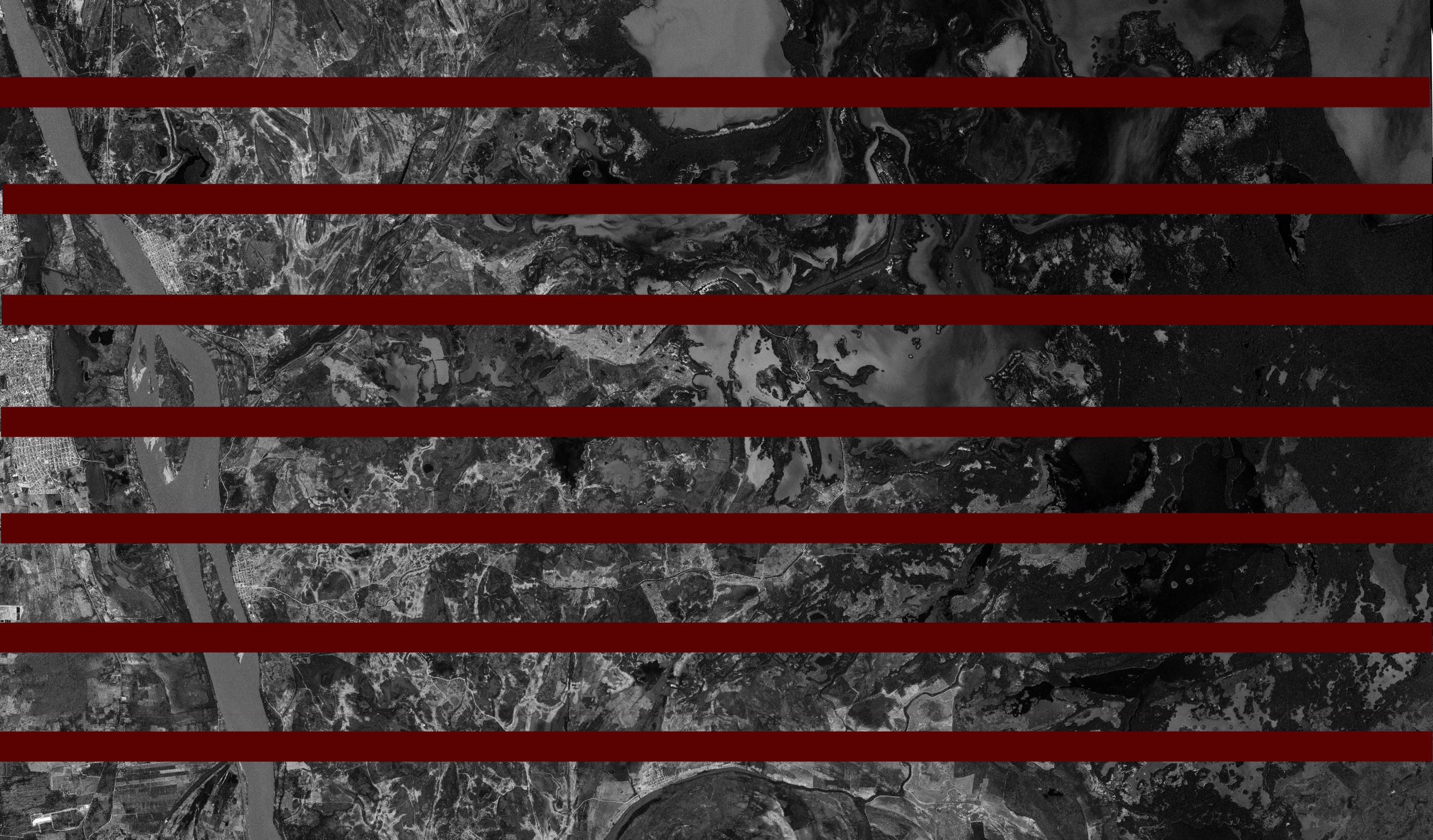} 
\caption{
Mock-up of the image from a single exposure of the camera on a SuperDove satellite. The image is 
divided into eight zones, one for each spectral band. This mock-up shows the whole scene in a single band, blue,
for simplicity whereas the images within the zones would be in different spectral bands. 
The red horizontal bars represent regions where the view is masked by the joints between
the eight individual filters assembled together preceding the sensor. 
This representation is based on publicly available information and may not represent a single-frame
image in detail.
}
\label{single_frame}
\end{figure*}

The second step requires some educated guesswork about the proprietary and undisclosed design and operation of the satellites.
The CCD sensor is preceded in the optical path by a filter with eight zones for the spectral bands. A single exposure
of the camera produces an image conceptually similar to figure \ref{single_frame}. For simplicity, figure 
\ref{single_frame} shows an image from satellite 247d in a single band, blue, whereas in the actual image, each of
the horizontal zones is the image through one of the eight different spectral filters. The opaque red bars represent regions
where the different filters are joined and the view through the filter is blocked. We consider the eight zones
on the sensor to be divided by the horizontal midpoints of the red bars. The image has been rotated as if the
satellite in a descending orbit were moving from north to south or vertically in the image.

The time required for the satellite to move the ground footprint of one sensor zone over the adjacent zone is $\sim 0.39$ s depending
on the altitude of the satellite. Ignoring the obscured regions,
if the camera frame rate were synchronized with the rate, $\sim 1/0.39 = 2.56$ s$^{-1}$, 
at which one zone moves over another
(the color-crossing rate),
a continuous mosaic in each of the colors could be constructed from eight exposures. However, to include the view of the ground
below the blocked regions of the filter, the camera frame rate needs to be faster.  
Also the color-crossing rate
is continuously increasing as the orbit of the satellite slowly decays.  Since synchronization is not required to construct the mosaic,
we assume that the camera frame rate is constant but asynchronous with the color-crossing rate.
From an analysis
of the images of the balloons, explained in the next section \S\ref{Verification}, we derive the camera frame rate as 
$\sim 1/0.18 = 5.55$ s$^{-1}$, 
just faster than twice the color-crossing rate. Thus, the mosaic in a single spectral band of an 
image with as many pixels as the sensor is constructed from
approximately 16 exposures of the camera. Importantly for our estimation of the apparent speed of an object, 
the difference in acquisition time between any two pixels of different colors
is either the color-crossing time, $0.39$ s, or the color-crossing time plus or minus the time between exposures  0.184  s
(estimated in \S\ref{Verification}).
The formula for the apparent velocity is then equation 2 of \citet{KW2023},
\begin{equation}
v_i = \frac {\Delta p} {\Delta t_{color} + a}
\label{eqn2}\end{equation}
where $\Delta p = p_i(x,y) - p_{i+1}(x,y)$ is the distance between the locations of the object in two color bands, $i$ and $i+1$. 
The variable $a$ can take one of three values:  $0, \pm \Delta  t_{camera}$. Equation \ref{eqn2} here is explained equivalently but  
differently in \citet{KW2023}.

Analysis of the images from satellite 2479, discussed in \S\ref{Verification} indicates that the mosaicing procedure is more complex 
than a simple translation of the full images to align with features on the ground.
Refraction due to atmospheric turbulence may distort regions of the image that can be corrected in the mosaicing. 
Microscale turbulence on time scales of microseconds typically results in blurring on arc second scales, approximately
the pixel size, and cannot be corrected by mosaicing. 
Less frequent mesocale turbulence with a longer spatial and time scale may shift the apparent
position of ground features within an image. Since most of the visual distortion results from the denser atmosphere
near the ground, the positional accuracy of objects at high altitude may be adversely affected by mosaicing corrections to align
ground features. The significance of this effect and
the circumstances in which it might apply are difficult to assess.  
The images from satellites 2439 and 247d do not appear to be affected.


\subsection{Verification}\label{Verification}

We apply the proposed method to the images
from satellites 2439 and 247d showing balloons over Missouri and Colombia, in figures 
\ref{positions2439} and \ref{positions247d}, respectively. Both
show equal displacements of the apparent position of the balloons in the different
spectral bands with three exceptions. Figure \ref{positions2439} shows that the displacement between the image of the 
balloon in the center of the figure with respect to the adjacent image of the balloon just to the north is about
half the length of the other displacements. Figure \ref{positions247d} shows two such occurrences at different places.
This indicates that the value of $a$ in equation \ref{eqn2} has a value of the
negative of the time delay between camera exposures of the color bands at these locations in the image and zero
elsewhere along the track of the balloon.

We can estimate the camera frame rate as follows. With the assumption that the balloon has no proper motion
and therefore no acceleration,
we expect equally spaced displacements between
color bands except for the change caused by the time delay between exposures. The time delay
between exposures is then found from equation \ref{eqn2} as the value that results in a constant velocity in the direction
of the satellite orbit. The time delay between exposures is then $0.184 \pm 0.007$ s for both balloons and the 
average apparent velocities are $338\pm 5.0$ and $371\pm 4.3$ m/s.
The errors are on the apparent velocities are estimated empirically from the standard deviations of the 
measurements in table \ref{velocity_table}.

\begin{table}[!h]
\caption{Displacements and Apparent Velocities of Balloon Images Between Color Bands}
\begin{tabular}{cc cc} 
\toprule
\multicolumn{2}{c} {2439}		& \multicolumn{2}{c} {247d} \\
\cline{1-2}				\cline{3-4}
Segment 	& App. Vel. 	& Segment	 &App. Vel. \\
(m) 		& (m/s) 		& (m) 		& (m/s) \\ 
\colrule
120		& 335		& 67			& 371 	\\
122		& 340		& 136		& 371	\\
120		& 335		& 136		& 371	\\	
 60		& 345		& 136		& 375	\\
120		& 335		& 68			& 375	\\
122		& 343		& 133		& 363	\\	
119		& 332		& 138		& 377	\\
\botrule
\end{tabular}
\label{velocity_table}
\end{table}

We can also estimate the altitude of a balloon from the ratio of the distance traveled by the
satellite to distance traveled by the image of the balloon on the ground,
\begin{equation}
\frac {h_{sat} - h_{ball}} {h_{ball}} = \frac{ L_{sat}} {L_{ground}}
\end{equation}
We find altitudes of  $21206 \pm 341$ m and $21505\pm 442$ m for the first and second balloons over
Missouri and Colombia respectively.
The altitude of the first balloon at a different position, over South Dakota,
was previously reported by Planet Labs as 20117 m (with no estimated error) as measured from
an image taken by satellite 242d. \citep{NYT2023}. The difference in the two reported altitudes is within 3$\sigma$.
It's also possible that the balloon was attempting to escape by fleeing to higher altitude during its eastward travel  over the US.
The altitude of the second balloon over Colombia has not previously been reported.

\begin{figure*} [!ht]
\begin{tabular} {ll}
\includegraphics[width=3.5in,trim={0.in 0.0in 2.5in 6.0in},clip]{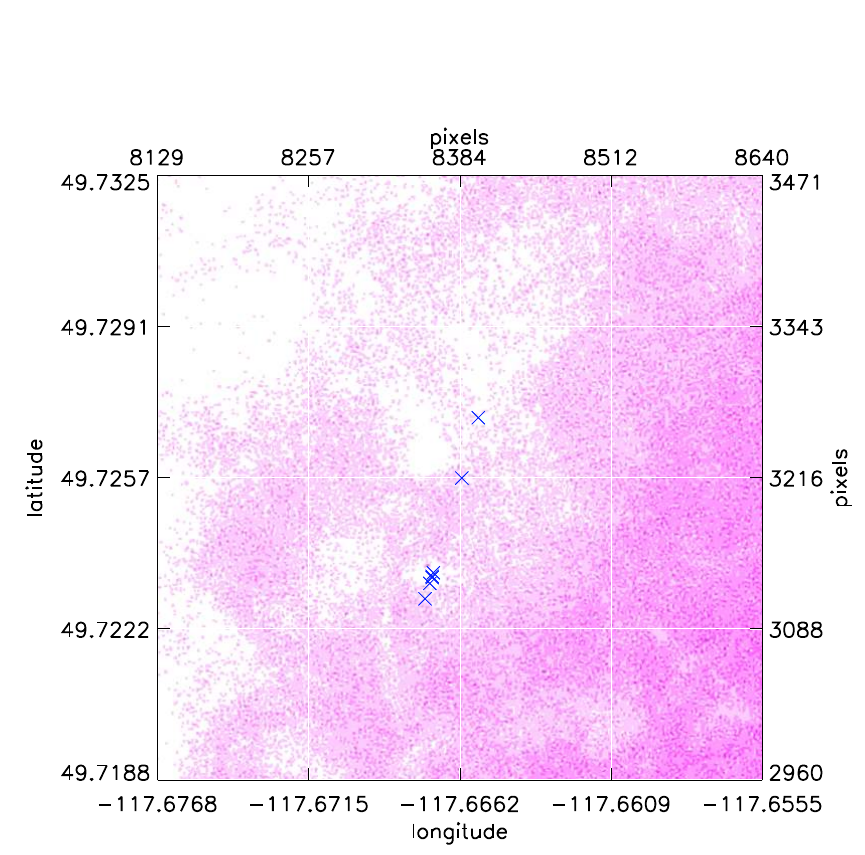} &
\includegraphics[width=3.5in,trim={0.in 0.0in 2.5in 6.0in},clip]{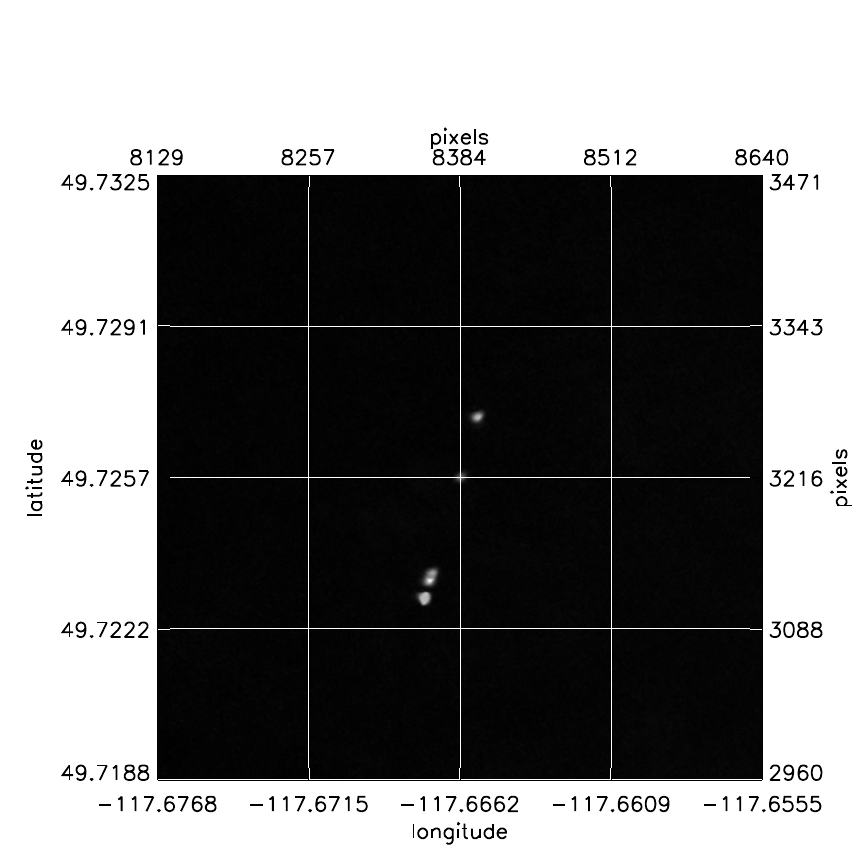} \\
\end{tabular}
\caption{
Positions of the first Chinese balloon over British Columbia in the eight bands of SuperDove satellite 2479 against a
3-color visual background ({\it left}) and against the summed absolute values of the seven differenced pairs of
images in the eight spectral bands ({\it right}) in the format of figure \ref{positions2439}. The pink color in the visual image in this
image from satellite 2479 appears to be due to severe digital compression of the archived image. The actual scene should be all white.}
\label{positions2479}
\end{figure*}

\subsection{Loss of Ground Referencing over Snow and Ice}

The first balloon was also spotted in an image made by SuperDove satellite 2479 while it was over British Columbia, Canada, before
it crossed into the US. Figure \ref{positions2479} shows the apparent positions of the balloon in the same format
as figures \ref{positions2439} and \ref{positions247d}. 
The ground is completely covered with white snow and ice consistent with the winter conditions, and no features for ground
reference that are required for positional accuracy are apparent. The only significant feature in any of the spectral bands is the balloon
itself. The mosacing algorithm applied by Planet Labs has evidently mistaken the exceptionally bright
balloon for a ground reference feature
and attempted to align the spectral images on the balloon itself. This results in physically impossible apparent locations
for the balloon.

\section*{Note Added}
This research is conducted as part of the Galileo Project  at Harvard University 
 whose goal is to collect
scientific quality data that may be useful in the search for objects of extraterrestrial origin
(https://projects.iq.harvard.edu/galileo/home).
\section*{Acknowledgments}
We acknowledge Planet Labs Corp. for access to the data and for technical support.


\bibliography{Verification}{}

\mycomment{

}

\end{document}